\documentclass[showpacs,preprintnumbers,amsmath,amssymb,prb,twocolumn,superscriptaddress]{revtex4}

\usepackage{graphicx}
\usepackage{dcolumn}
\usepackage{bm}
\def\comment#1{}

\def\slashchar#1{\setbox0=\hbox{$#1$}           
   \dimen0=\wd0                                 
   \setbox1=\hbox{/} \dimen1=\wd1               
   \ifdim\dimen0>\dimen1                        
      \rlap{\hbox to \dimen0{\hfil/\hfil}}      
      #1                                        
   \else                                        
      \rlap{\hbox to \dimen1{\hfil$#1$\hfil}}   
      /                                         
   \fi}                                         %

\newcommand{\hh}{\begin{picture}(13,9)(-2,2)
	\put (0,0) {\line (1,0) {8}}
	\put (8,8) {\line (-1,0) {8}}
	\put (0,0) {\circle*{3}}
	\put (0,8) {\circle*{3}}
	\put (8,0) {\circle*{3}}
	\put (8,8) {\circle*{3}}
	\end{picture}
}
\newcommand{\vv}{\begin{picture}(13,9)(-2,2)
	\put (0,0) {\line (0,1) {8}}
	\put (8,8) {\line (0,-1) {8}}
	\put (0,0) {\circle*{3}}
	\put (0,8) {\circle*{3}}
	\put (8,0) {\circle*{3}}
	\put (8,8) {\circle*{3}}
	\end{picture}
}

\begin{document}

\title{Renormalization, duality, and phase transitions in two- and three-dimensional quantum dimer models}

\author{Flavio S. Nogueira}
\email{nogueira@physik.fu-berlin.de}
\affiliation{Institut f{\"u}r Theoretische Physik,
Freie Universit{\"a}t Berlin, Arnimallee 14, D-14195 Berlin, Germany}

\author{Zohar Nussinov}
\email{zohar@wuphys.wustl.edu}
\affiliation{Physics Department, CB 1105, 
Washington University, 1 Brookings Drive, St. Louis, MO 63130-4899}

\date{Received \today}

\begin{abstract}
We derive an extended lattice gauge theory type action for quantum dimer models and relate it
to the height representations of these systems. We examine the system in two and three
dimensions and analyze the phase structure in terms of effective theories and duality arguments. 
For the two-dimensional case we derive the effective potential both at zero and finite temperature. 
The zero-temperature theory at the  Rokhsar-Kivelson (RK) point 
has a critical point related to the self-dual point of a class of $Z_N$ models in the $N\to\infty$ limit. 
Two phase transitions featuring a fixed line are shown to appear in the phase diagram, one at zero temperature and at the 
RK point and another one at finite temperature above the RK point. The latter will be shown to correspond to a Kosterlitz-Thouless (KT) 
phase transition, while the former will be governed by a KT-like universality class, i.e., sharing many features with a KT transition but 
actually corresponding to a different universality class.  
On the other hand, we show that 
at the RK point no phase transition happens at finite temperature.  
For the three-dimensional case we derive the corresponding dual gauge theory model at the RK point. We show in this case that at zero temperature a first-order phase 
transition occurs, while at finite temperatures both first- and second-order phase transitions are possible, depending on the relative values of 
the couplings involved.  
\end{abstract}

\maketitle

\section{Introduction}

The ``quantum dimer model" (QDM)\cite{RK, Review}  was introduced to emulate the quintessential features 
of valence bond states in spin systems. Its inception was motivated by 
the short-range resonating valence-bond (RVB) state of Anderson,\cite{Anderson-1973}
which was used as a possible starting point for a theory of high-$T_c$ superconductors.\cite{Anderson-1987} 
Since, the QDM has played a prominent role in modeling 
various frustrated magnets,\cite{Review} cold atom systems,\cite{buchler}  
Josephson junction arrays,\cite{albuq} spin-orbital systems,\cite{vernay} 
topological quantum orders,\cite{Wen_book} fractionalization, and other 
phenomena.\cite{Review} The only degrees of freedom in the QDM 
are dimers that represent singlet 
states formed by neighboring spins on the lattice. 
Two spin singlet states cannot overlap or share
a common lattice site. Similarly, any two dimers within the QDM 
cannot overlap--- the dimers satisfy a ``hard core'' constraint. 
Several spin \cite{MG, klein, Nussinov-2007, uniqueness}, and orbital models 
\cite{ishihara}
indeed have ground states
which are precisely of the dimer type.
In several spin systems, pairs of $S=1/2$ spins bind into singlet states.
In some spin systems, such as the Klein models, \cite{klein}
it can be proven that not only are dimer states ground
states but that they are the only ground states on general lattices. \cite{uniqueness}

One of the major complications of real spin singlet systems
by comparison to the QDM, 
is the non-orthonormality of the singlet product basis states
(an item which should not be confused with viable
linear independence of these states).
In spin systems with dimer ground states is possible to systematically write down rules
for the evaluation of the overlap between 
the singlet states on bipartite \cite{RK, earlierRVB, Fradkin-Book}
and more general
lattices \cite{Nussinov-2007} as well as to evaluate the
general matrix elements of spin exchange and
other terms. \cite{Nussinov-2007, beach}
The QDM avoids many complications
by focusing on the quintessential physics
of hard core dimer systems. The overlap 
between different dimer states is simply set 
to zero.

The Hamiltonian of the QDM on a square lattice 
reads \cite{RK}
\begin{eqnarray}
H&=&\sum_{\Box}
\left[
-t\left(\left|\vv\right\rangle\left\langle\hh\right| + \left|\hh\right\rangle\left\langle\vv
\right|\right)\right.
\nonumber\\
&+&\left.v\left(\left|\vv\right\rangle\left\langle\vv\right|+\left|\hh\right\rangle\left\langle\hh
\right|\right)
\right],
\label{QDM}
\end{eqnarray}
with the sum performed over all elementary plaquettes of the square lattice. 
We reiterate that the dimer states in this model are {\em orthonormal},\cite{RK} differently from the 
singlet valence-bond states originary from quantum spin models, which are in general not 
orthogonal.\cite{Review} 

As seen in Eq.(\ref{QDM}), the QDM contains both a kinetic ($t$) term that flips one dimer tiling 
of any plaquette to another
(a horizontal covering to a vertical one and vice versa)
and a potential $(v$) term counting how many plaquettes are flippable.
These two terms --- kinetic and potential --- were treated
on different footing in earlier works. Particularly 
important is the point $t=v$, 
the so-called Rokhsar-Kivelson (RK) point,
where both the kinetic and potential terms are
of equal magnitude. The RK point is 
an exactly solvable critical point 
that separates two different 
valence-bond solid (VBS) phases. 

Within the VBS phases, the dimers  
break lattice point group symmetries.\cite{Review, Syljuasen} 
On non-bipartite lattices and $t\neq v$, valence 
bond liquid phases appear. On the triangular lattice,\cite{MSl} short ranged 
resonating valence bond phase with 
no gapless excitations and with deconfined, gapped, spinons 
appear for a finite range of parameters. Similar behavior is 
also found on the kagome lattice.\cite{MSP} For square lattices, valence-bond 
liquid phases seem to be possible only at the RK point.

It has been shown in several 
recent papers\cite{Syljuasen,Castelnovo,Misguich-2008,Ralko-2008,Charrier,Chen} that the phase structure of the QDM is 
considerably richer than previously thought. Especially interesting here is the emergence and characterization of the 
different VBS phases. In particular, it has been recently shown \cite{Ralko-2008} that the VBS phase continuously interpolates 
with the plaquette phase via a mixed regime.\cite{Leung} Classical 
three-dimensional dimer models have also recently shown to exhibit interesting phase transitions.\cite{Charrier,Chen} Similar to some quantum phase 
transitions in antiferromagnets \cite{science,NKS} several 
three-dimensional classical dimer models seem to exhibit second-order phase transitions that do not easily fit within a Landau-Ginzburg-Wilson 
approach.\cite{Chen} 

In this paper we will study the QDM  of Eq.(\ref{QDM}) via effective field theories and duality methods. 
Our work will focus on the square and cubic lattices.  An important goal of this work is to elucidate the nature of the phase transition, both at zero and at finite temperatures. We characterize the 
phase transitions in two and three dimensions by employing an interplay between renormalization 
and duality. The plan of the paper is as follows. In Section \ref{spin_rep}, we will
introduce a simple spin $S=1/2$ representation of the QDM which
will allow us in Section \ref{gauge_rep} to introduce a new lattice gauge theory representation of the QDM. We show that the QDM is equivalent to an extended Abelian lattice gauge theory in which in addition to the usual plaquette cosine term of standard lattice gauge theories, a higher harmonic of the field strength
is also present. It is known from the lattice gauge theory literature \cite{Bhanot} that in four spacetime dimensions such a theory has a rich phase 
structure. In Section \ref{height-model}, the two-dimensional QDM model is discussed using an effective height model. 
Such height model representations of the QDM are well known in the literature,
\cite{Subir,Fradkin-2004,Henley,Review,Zeng-Henley} and have been motivated by
physical arguments. Here, we put these representations in perspective via the 
extended lattice gauge theory that we derive 
in Section \ref{gauge_rep}. We show that the height field theory model can be derived from the 
extended lattice gauge theory. It should be stressed here that the extended lattice gauge theory 
in Section II is more precise than the height models discussed in Section \ref{height-model}. However, studying it by purely analytical means is very 
difficult, so that effective height models are indeed useful in this respect. In Section \ref{transitions}, the phase structure of the two dimensional QDM  is analyzed using 
effective potentials and duality arguments. The phase diagram is discussed both at zero and finite temperatures. We show that 
the full phase diagram features two Kosterlitz-Thouless (KT) like phase transitions, one at zero temperature at the RK point, and another one 
at finite temperature away from the RK point at $t>v$. The latter is a genuine KT transition, while the former is KT-like, i.e., it shares many 
properties with the usual KT transition, but it actually corresponds to a new universality class featuring a fixed line. 
We further illustrate that the effective potentials may be the same in both situations. In Section \ref{3Dsection},
 we discuss the three-dimensional QDM at the RK point at both zero and finite temperatures. To this end, the extended lattice gauge theory is considered more directly in a Villain approximation. Here duality plays a crucial role in 
determining the phase structure of the 
RK point.  We show that at zero temperature and at the RK point a first-order phase transition between a 
VBS and a RVB state takes place. Thus, in contrast with the two-dimensional QDM, {\em no quantum critical point} exists in three dimensions at the RK point. 
For finite temperatures both first- and second-order phase transitions are possible.
The character of the transition 
(whether it is continuous or
abrupt) depends on the relative values of the couplings involved.  We summarize our conclusions
in Section \ref{endsection}.

\section{Spin representation of the QDM}
\label{spin_rep}

In this section, we will find a direct representation for the QDM in terms of a lattice gauge theory.
This will allow us to systematically derive the height representation as an 
approximation.  To achieve this aim, we first express the QDM directly
in a spin language. As we will show in a future publication, the
approach that we introduce below enables a derivation of spin representations
and ensuing gauge type theories
for other lattices with general (non-square type) elementary plaquettes.
We will employ a
simple algebraic property of the QDM:
 {\em The potential term in Eq.(\ref{QDM}) is the square of the kinetic term},
\begin{eqnarray}
\left(\left|\vv\right\rangle\left\langle\hh\right| + \left|\hh\right\rangle\left\langle\vv
\right|\right)^{2} \nonumber
\\ = \left|\vv\right\rangle\left\langle\vv\right|+\left|\hh\right\rangle\left\langle\hh
\right|.
\label{square}
\end{eqnarray}
What enables this relation is the orthonormality of the dimer states. \cite{RK}
The relation of Eq.(\ref{square}) will enable us to treat both the kinetic and potential terms within a uniform systematic framework later on. All (positive) even powers of the kinetic term
give rise to the potential energy
term whereas all odd powers of the kinetic energy term yield the kinetic term
unchanged.
Denoting the kinetic term by $B_{\Box}$, 
Eq.(\ref{QDM}) can be rewritten as
\begin{eqnarray}
H = \sum_{\Box}(-tB_{\Box} + v  B_{\Box}^{2}),
\label{HB}
\end{eqnarray}
Thus, for a given value of $v>0$, the ground states
minimize the sum $\sum_{\Box}(B_{\Box} - t/(2v))^{2}$
over all plaquettes of the latttice. 
As will become evident later on, the kinetic term $B_{\Box}$ will play,
in the gauge representation, a role similar to that of a (modular) magnetic
flux that threads a plaquette. The flux pattern $B_{\Box}$ within the 
ground states will be determined by the ratio $t/v$. 

We will now employ and cast the relation of Eq.(\ref{square}) in a spin 
language. Similar to Refs. \onlinecite{Fradkin-Book}, \onlinecite{Orland}, and \onlinecite{Moessner-2001},  
we designate the presence/absence of a dimer between 
the two sites $i$ and $j$ by $\sigma^{z}_{ij} = 1$ and $\sigma^{z}_{ij}=-1$ respectively.  
Introducing the Pauli raising/lowering operators
 $\sigma_{ij}^{\pm} = \frac{1}{2} (\sigma_{ij}^{x} \pm i \sigma_{ij}^{y})$,
the QDM Hamiltonian reads

\begin{eqnarray}
H = - t \sum_{\Box} (W_\Box+W^{\dagger}_{\Box}) + v \sum_{\Box} (W_\Box W_\Box^{\dagger}
+ W_{\Box}^{\dagger} W_{\Box}), 
\label{rk1/2}
\end{eqnarray}
where 
\begin{eqnarray}
W_{\Box} = \sigma_{ij}^{+} \sigma_{jk}^{-} \sigma_{kl}^{+} \sigma_{li}^{-},
\label{W_defn}
\end{eqnarray}
and $\Box= ijkl$ is a plaquette. The spin representation of the kinetic only term was discussed 
in Refs. \onlinecite{Orland, Moessner-2001, Henley}. Our new full Hamiltonian of Eq. (\ref{rk1/2}),
containing both the kinetic and potential terms, is
a faithful representation of the QDM of Eq. (\ref{QDM}). It automatically
incorporates the hard-core constraint of the QDM that prevents two dimers from
overlapping. If we denote the two
vertical bonds of the plaquette $\left|\vv\right\rangle$, by $(ij)$ and $(kl)$, 
then $W_{\Box}^{\dagger} \left|\vv\right\rangle = \left| \hh\right\rangle$. 
For any state $|\psi \rangle$ orthogonal to $\left|\vv\right\rangle$, we have
$W_{\Box}^{\dagger}|\psi \rangle =0$. Similarly, 
$W_{\Box} \left| \hh\right\rangle  =  \left|\vv\right\rangle $. 
The products
$(W_\Box W_\Box^{\dagger})$ and $(W_\Box^{\dagger} W_\Box)$
constitute projection operators 
onto the states $\left|\vv\right\rangle$ and $\left|\hh\right\rangle$
respectively. As $(\sigma_{ij}^{+})^{2} = (\sigma_{ij}^{-})^{2} = 0$,
we obtain the Hamiltonian of Eq.(\ref{HB})
where
\begin{eqnarray}
B_{\Box} = W_{\Box} + W_{\Box}^{\dagger}. 
\label{BWW}
\end{eqnarray} 
The kinetic term in Eq. (\ref{rk1/2}) is precisely the ring exchange term in a well studied model (in its
XY version) \cite{sandvik} in which we regard the centers of bonds of a square lattice as vertices of a square 
lattice rotated by 45 degrees and a scaled down lattice constant by a factor of $2^{-1/2}$. 
In the large spin limit, 
Eq.(\ref{rk1/2}) is a classical XY Hamiltonian.
By introducing a gauge field representation of the spins, we can check whether deconfined 
criticality  \cite{science,NKS} may arise.

\section{Lattice gauge theory representation}
\label{gauge_rep}

We will now transform the exact spin representation into a lattice
gauge theory by writing down the spin coherent basis action corresponding to the 
Hamiltonian of Eq. (\ref{HB}). To this end, the spin $({\bf{s}}= \frac{1}{2} {\bf{\sigma}})$
points anywhere on a sphere of radius $s=1/2$. In imaginary time,
the spin performs a cyclic evolution  on the spin during the time 
interval $0 \le \tau < \beta$. 
The Euclidean action
is 
\begin{eqnarray}
S= - is \sum_{ij} (2 \pi - {\cal{A}}_{ij}) +  \int_{0}^{\beta} d \tau~ H[\{{\bf{ \sigma_{ij}}}\}],
\label{wzws}
\end{eqnarray}
where in the first (Wess-Zumino-Witten (WZW) borne) term, ${\cal{A}}_{ij}$ is the area subtended by the spin ${\bf{s}}_{ij} = \frac{1}{2} {\bf{\sigma}}_{ij}$ as it cyclically 
evolves in imaginary time ($ 0 \le \tau < \beta$) on the sphere. This
term corresponds to the Berry phase within the coherent spin basis.
In Eq.(\ref{wzws}),  $\{{\bf{s}}_{ij}\}$ denotes all spins of the lattice
(all bond centers). 

Parameterizing the spin on
the bond $(ij)$ by its location on 
the sphere by
${\bf{s}}_{ij} = s (\sin \theta_{ij} \cos A_{ij}, \sin \theta_{ij} \sin A_{ij}, \cos \theta_{ij})$,
the Hamiltonian of Eq. (\ref{HB}) reads
\begin{eqnarray}
 H=2 \sum_{i,m,n} [-t (\sin \theta)_\Box \cos(F_{imn})\nonumber
 \\ +v  (\sin \theta)_\Box^{2} \cos(2F_{imn}) ],
 \label{Hxyl}
\end{eqnarray}
where $F_{imn}=\nabla_m A_{in}-\nabla_n A_{im}$ is the field strength in the lattice, with the latin 
indices $m,n$ being summed over $(d-1)$ spatial dimensions, and $i$ a site label.  The factor 
$(\sin\theta)_\Box$ denotes the product of four factors of $\sin\theta_{ij}$ around a plaquette. 
The area subtended by the spin on the sphere depends on $\theta_{ij}$.
The WZW contribution can be rewritten as 
\begin{eqnarray}
s \int_{0}^{\beta}  d\tau~\left(i \cos \theta_{ij}  \frac{d A_{ij}}{d \tau}\right)
\label{area}
\end{eqnarray}
The height representation of the QDM was introduced and motivated
by numerous insightful intuitive considerations in earlier works 
\cite{Review, Moessner-2001, Subir, Zeng-Henley, Fradkin-2004,Vish-2004}.
In the present work, we derive it by a sequence of 
approximations. 

The partition function
\begin{eqnarray}
Z = \int \prod_{ij} d \theta_{ij} ~dA_{ij} e^{-S},
\end{eqnarray}
with $S$ the action of Eq.(\ref{wzws}). 
Henceforth, we will make several approximations.
In the Hamiltonian, we will replace 
the explicit angular dependence on the angles $\theta_{ij}$
(to be integrated over) with an average value determined (within the coherent path
integral formulation) by setting $|s_{ij}^{z}| = 1/2$. Later on, we will 
further invoke both the continuum and Villain
approximations.

Within the coherent path integral formulation, $s_{ij}^{z}$ is continuous.
We approximate the integral over the polar angles $\theta_{ij}$ 
by enforcing the $S=1/2$ results that would have been obtained in the canonical
formulation (i.e. that would have been obtained for the averages of these
squares alone with the coherent spin path integral). To this end,
we replace in the Hamiltonian any appearance of 
$s^{2} \sin^{2} \theta_{ij} $ by  $1/2$, as the squared norm of the XY (planar) part of
the $S=1/2$ spin is $s(s+1) - (s^{z})^{2} = 1/2$. That is, in all pertinent expressions
for the spin $S=1/2$ system  we set $|s_{ij}^{z}| = 1/2$ and
\begin{eqnarray}
\sigma_{ij}^\pm= \frac{e^{\pm iA_{ij}}}{\sqrt{2}},
\label{fixsz}
\end{eqnarray}
with real compact $A_{ij}$. 
Substituting  this form in Eq. (\ref{W_defn}), 
we obtain, from Eq. (\ref{HB}), the Hamiltonian of an extended lattice gauge theory, 
\begin{equation}
 H=\sum_{i,m,n}\left[-\frac{t}{2}\cos(F_{imn})+\frac{v}{8}\cos(2F_{imn})\right].
 \label{Hxy}
\end{equation}

We can construct an action in which $\sigma_{ij}^z$ plays the role of a ``momentum'' that is conjugate to the phase $A_{ij}$,
\begin{equation}
 S=is\sum_{\tau,j,n}\sigma_{jn}^z\nabla_\tau A_{jn}+H.
 \label{area_conjugate}
\end{equation}
In Eq.(\ref{area_conjugate}), we discretize the imaginary time $\tau$.
The first term with $s \sigma_{jn} ^{z} \to s^{z}_{jn}$ is the WZW term of Eq. (\ref{area}) --- 
it is determined by the area ${\cal{A}}$ swept by the spin ${\bf{s}}_{jn}$ as it
traverses the sphere in the closed orbit for $0 \le \tau < \beta$ [see Eq. (\ref{area})]. 
 This Berry phase is similar to the one introduced in the Sachdev-Jalabert model of quantum 
antiferromagnets.\cite{Sachdev-Jalabert}  
The WZW term may also be rederived from another vista.
As $[\sigma_{ij}^z,\sigma_{ij}^{\pm}]=\pm 2\sigma_{ij}^\pm$, it 
follows that the lattice gauge field $A_{ij}$ is canonically conjugate to $\sigma_{ij}^z$ (this is similar to the discussion in Ref. \onlinecite{NB} in 
a different context). 
%
The partition function is therefore given by
\begin{equation}
 Z=\int \prod_{in} dA_{in}\sum_{\{\sigma_{in}^z=\pm 1\}}\exp\left(-is\sum_{\tau,j,n}\sigma_{jn}^z\nabla_\tau A_{jn}-H\right).
\end{equation}
As stated earlier, within the coherent path integral formulation,
$s_{ij}^{z}$ is continuous. To be consistent, we invoke the same
approximation that we employed earlier 
($|s_{ij}^{z}| =1/2$) and set $\sigma^{z} = \pm 1$.
Summing over $\sigma^{z}_{ij}$, this 
leads, up to an irrelevant constant, to the result
\begin{eqnarray}
 Z&=&\int \prod_{in} dA_{in}\prod_{\tau,j,n}\cos(s\nabla_\tau A_{jn})\exp\left(-H\right)\nonumber\\
&=&\int \prod_{in} dA_{in}~e^{-\tilde S},
\end{eqnarray}
where the action in terms of the gauge field only is given by  
\begin{eqnarray}
\label{Action}
 \tilde S&=& -\sum_{i,\tau,n}\ln\left[\cos(s \nabla_\tau A_{in})\right]\nonumber\\
&+&\sum_{i,\tau,m,n}  \left[ -\frac{t}{2}\cos(F_{imn})+\frac{v}{8}\cos(2F_{imn})\right].
\end{eqnarray}
A related theory in which the first term in Eq. (\ref{Action}) was absent was studied numerically 
in four dimensions.\cite{Bhanot} Such a theory usually has a rich phase structure 
characterized by a tricritical point. 
A classical XY version was also studied in three dimensions.\cite{Janke-Kleinert}
Note that the extended lattice gauge action of Eq. (\ref{Action}) differs from 
earlier proposals for the QDM.\cite{Fradkin-Book, Moessner-2001}
Amongst other benefits, our lattice gauge action enables us 
to derive the height model as an approximation without resorting
to phenomenological arguments. It also allows us to immediately
identify $t=v$ as a special point. 

Expanding the action (\ref{Action}) up to quadratic order in the fields, taking the 
continuum limit, and rescaling both $t$ and $v$ by a uniform
factor, $t \to s^{2} t, ~ v \to s^{2}v$, we obtain up
to an overall innocuous multiplicative factor, a Lagrangian 
density of the form, 
\begin{equation}
 \tilde L\approx \frac{1}{2}(\partial_\tau {\bf A})^2+\frac{t-v}{4}F_{mn}^2.
\label{gauge_theory-0}
\end{equation}
When we use our quadratic gauge action,
we have a dependnce only on $(t-v)$
and not on the individual values of $t$
and $v$. This
symmetry in the quadratic order gauge (and, as we
will see later on, the derived height model)
description of the QDM  is
lifted as higher order terms
are included in our gauge action.

By analytically continuing to real time, we have the following gauge field propagator in the transverse gauge $\nabla\cdot{\bf A}=0$,
\begin{equation}
 D_{mn}(\omega,{\bf p})=\frac{1}{\omega^2-(t-v){\bf p}^2}\left(\delta_{mn}-\frac{p_mp_n}{{\bf p}^2}\right).
\end{equation}
We see that for $t>v$ the spectrum of elementary excitations is given by
\begin{equation}
\label{phonon}
\omega=\sqrt{t-v}|{\bf p}|.
\end{equation}
For $t<v$, on the other hand, we have a purely damped mode. This is not a completely satisfactory physical result, since it makes the 
system unstable, especially at finite temperature. In order to stabilize the system in this case, we have to include higher order 
terms, which can be obtained from the photon self-energy for $\omega$ small.

\section{Derivation of the two-dimensional height model}
\label{height-model}

The quintessential low-energy properties of 
the QDM can be captured by a height representation \cite{Review, Moessner-2001, Subir, Zeng-Henley, Fradkin-2004,Vish-2004}  where 
the weights (heights) $h$ are associated with the plaquettes of the 
underlying lattice. Our gauge theory action 
of Eq. (\ref{Action}) ---
an average coarse grained form of the exact QDM Hamiltonian in the form of Eq. (\ref{rk1/2}) ---
offers a new route towards systematically obtaining the height model representations
of the QDM. In this section, we will show to obtain these representations from the 
gauge theory of the previous section. 
The advantage of the lattice gauge theory approach is that it allows to easily obtain the field theories 
in both the two- and three-dimensional cases.
In a nutshell, the gauge theory and height representation are related by a simple (``Abelian-Higgs'' type) duality.
This duality amounts to the following rule of thumb in 2+1 dimensions,
\begin{eqnarray}
\epsilon_{\mu \nu \lambda} \partial_{\nu} A_{\lambda} \propto \partial_{\mu} h,
\label{Abelian_Higgs}
\end{eqnarray}
with $h$ the height field and where the greek indices run over the spacetime coordinates. 
That is, an effective magnetic field derived from our
gauge potential is equal to the gradients of the height field. Note that due to the compactness of the gauge field, 
the magnetic flux through a {\it closed} surface should be quantized.

To make it more lucid, we will briefly describe the physical content of the correspondence of 
Eq. (\ref{Abelian_Higgs}) for the cases represented in Fig. \ref{Fig:RK-phases}.
We simply focus on the relation of Eq. (\ref{Abelian_Higgs})
to see to what height phases the various VBS phases correspond to.
First, we define the flux per plaquette $\Phi$ as the counter-clockwise sum of the
(directed) link variables $A_{ij}$.

(a) Inspecting Eq. (\ref{fixsz}), we see that when $\Phi= 0$ (mod $2  \pi$), 
no flux pierces the plaquettes and by Eq.(\ref{Abelian_Higgs}),  the height field $h$ is
a constant. This case corresponds to the so called plaquette phase, where the dimers resonate around 
a plaquette in a would-be columnar pattern.

(b) Similarly, the staggered VBS corresponds to 
half a flux quantum per plaquette
\begin{eqnarray}
\Phi = \pi ~~~~~~~~({\rm mod}~ 2 \pi).
\label{fluxs}
\end{eqnarray}
With a fixed circulation direction
flux in Eq. (\ref{fluxs}), this flux is seen to be
staggered
on the two sublattices for a fixed value 
of $A_{ij}$. In the case of $A_{ij} =\pi ({\rm mod}~ 2 \pi)$
as $\pi \equiv - \pi ({\rm mod}~ 2 \pi)$, a staggering of half-fluxons
is equivalent to a uniform
array of half-fluxons (sans staggering).

Both states (a) and (b) are 
invariant under the time reversal
operation which corresponds to
$A_{ij} \to -A_{ij}$.

%
%

To derive the correspondence of Eq. (\ref{Abelian_Higgs})
we will follow a sequence of steps below. (i) We will write down a Villain type action
that reduces to Eq. (\ref{gauge_theory-0}) in a continuum limit.
(ii) We will then apply the Poisson summation twice on this action.
(iii) Finally, we will show that the continuum limit of (ii) leads to
a height model with the substitution of Eq.(\ref{Abelian_Higgs}).

(i)  We start by introducing the following effective lattice action in the Villain approximation:

\begin{eqnarray}
 \label{gauge_theory-0-new}
S&=&\frac{1}{2}\sum_{i,j}D({\bf x}_i-{\bf x}_j)
\nonumber\\
&\times&(\nabla_\tau A_{im}-2\pi L_{im})(\nabla_\tau A_{jm}-2\pi L_{jm})
\nonumber\\
&+&\sum_i\frac{1}{4c}(F_{imn}-2\pi N_{imn})^2,
\end{eqnarray}
where
\begin{equation}
D({\bf x}_i-{\bf x}_j)=\frac{1}{L}\sum_{{\bf p}}e^{i{\bf p}\cdot({\bf x}_i-{\bf x}_j)}D({\bf p}),
\end{equation}
with 

\begin{equation}
 D({\bf p})=\frac{1}{K^{-1}\left[4-2(\cos p_1+\cos p_2)\right]+\rho},
\end{equation}
where $L_{im}$, $N_{imn}$ are integer fields and $\rho\propto t-v$.
The parameter $c$ which will appear in the height model
that we will derive is usually set to unity in the literature. 
However, this should be avoided, as $c$ is not dimensionless. In fact, as we will 
see, it plays a crucial role 
in the characterization of the quantum phase transition of the theory. 
%
%

The Villain gauge theory action (\ref{gauge_theory-0-new}) constitutes a version of the original lattice gauge theory (\ref{Action}) of the 
QDM. Like all Villain actions, it has the advantage of being more tractable, and yet including all relevant physics of the problem.

Note that by considering a naive continuum limit and 
low momentum such that $D({\bf x}_i-{\bf x}_j) \approx \delta_{ij}/\rho$ and rescaling the gauge field as $A_m\to\sqrt{\rho}A_m$,  with $\rho\propto (t-v)$, we obtain 
a Lagrangian of the form (\ref{gauge_theory-0}).
This low-momentum behavior reflects the fact that neglecting the compactness of the gauge field 
in Eq. (\ref{gauge_theory-0-new}), a Bogoliubov-like spectrum is obtained,  

\begin{equation}
\label{Bog-spec}
 \omega=\sqrt{\frac{\rho}{c}p^2+\frac{p^4}{cK}}.
\end{equation}
For $\rho>0$ and $p$ small we approximately recover the spectrum (\ref{phonon}). We see that for $\rho<0$ a momentum space instability occurs. 
We will come back to this point later.

(ii)  We next dualize the action of Eq.(\ref{gauge_theory-0-new}). By using the Poisson summation formula we rewrite the action in terms of auxiliary integer fields $M_{j0}$ and $M_{jm}$ in the 
form

\begin{eqnarray}
 S&=&\sum_j\left[\frac{c}{2}M_{j0}^2+\frac{\rho}{2}{\bf M}_j^2+\frac{1}{2K}(\nabla{\bf M}_j)^2\right]
\nonumber\\
&-&i\sum_j(M_{j0}\epsilon_{0mn}\nabla_mA_{jn}+M_{jm}\epsilon_{m0n}\nabla_\tau A_{jn}).
\end{eqnarray}
We now use partial summation and integrate out the compact gauge field to obtain the constraint

\begin{equation}
 \epsilon_{\mu\nu\lambda}\nabla_\nu M_{j\lambda}=0.
\end{equation}
The constraint is solved by 

\begin{equation}
 M_{j0}=\nabla_\tau N_j,~~~~~~~~~{\bf M}_j=\nabla N_j,
\end{equation}
where $N_j$ is an integer field. This leads to the action

\begin{equation}
 S=\frac{1}{2}\sum_j\left[c(\nabla_\tau N_j)^2+\rho(\nabla N_j)^2+\frac{1}{K}(\nabla^2N_j)^2\right].
\end{equation}
By using the Poisson formula once more to convert the integer field $N_j$ into a real field $h_j$, we obtain

\begin{eqnarray}
\label{Action-0}
 S&=&\frac{1}{2}\sum_j\left[c(\nabla_\tau h_j)^2+\rho(\nabla h_j)^2+\frac{1}{K}(\nabla^2h_j)^2
\right.\nonumber\\
&-&\left.2\pi i~ n_jh_j\right],
\end{eqnarray}
where $n_j$ is a new integer field.

(iii)  In the continuum limit of the action (\ref{Action-0}) 
the ``charges'' $n_j=\pm 1$ are the most relevant ones, so that in the 
grand-canonical partition function the (product in $j$) of factors 
$\exp(i2\pi n_jh_j)$ exactly exponentiates,\cite{Polyakov} producing 
a term $\cos(2\pi h)$ in the continuum limit of the action.  
Thus, we now finally arrive at a directly 
derived form the imaginary time Lagrangian
of the height model

\begin{equation}
\label{L}
 {\cal L}=\frac{c}{2}(\partial_\tau h)^2+\frac{\rho}{2}(\nabla h)^2+\frac{1}{2K}(\nabla^2 h)^2-z\cos(2\pi h).
\end{equation}

Comparing Eq.(\ref{L}) with Eq.(\ref{gauge_theory-0-new}), the 
correspondence of Eq.(\ref{Abelian_Higgs}) becomes manifest. 
The height model of Eq.(\ref{L}) has been introduced and motivated
by numerous insightful intuitive considerations in earlier works. 
\cite{Review, Moessner-2001, Subir, Zeng-Henley, Fradkin-2004,Vish-2004}
Our approach enables a direct derivation
of the height model from the original QDM Hamiltonian of Eq.(\ref{QDM}).

Putting all of the pieces together, in 2+1 dimensions, our lattice gauge theory
can be rewritten as a Lagrangian of Eq.(\ref{L}) for a scalar field, 
with the term $-z\cos(2\pi h)$ following from 
the compactness of the lattice gauge field.\cite{Polyakov}  

In the case of 3+1 dimensions which we will dwell on later, the dual field strength is a second 
rank tensor, just as the original field strength. Therefore, we cannot introduce a scalar field in this case and have to work further with a 
gauge field. We will consider this case later on in detail (Sect. IV).

The action of Eq.(\ref{Action-0}) may, in principle, also lead to higher cosine harmonics, like for example $\cos(2\pi mh)$ with $m=2,3,\dots$. However, 
as long as the couplings in front of all factors $-\cos(2\pi mh)$ are positive, the higher harmonics are irrelevant in the Renormalization Group sense.\cite{ZJ} 
On the other hand, there are situations where couplings with a negative sign play an important role, like in the case of emerging 
plaquette and mixed phases, as discussed recently in Ref. \onlinecite{Ralko-2008}.

\section{Transitions in the two-dimensional system}
\label{transitions}

In this section, we study the structure of the phase diagram 
of the height model of Eq. (\ref{L}). The outcome of our analysis
is summarized in Table I. 

\begin{table}
\caption{Summary of the phase structure of the two-dimensional height representation
of the QDM for different temperatures $T$ and $\rho \propto (t-v)$. 
Kosterlitz-Thouless (KT) type transitions occur when (i)
$T=\rho =0$ or (ii) $T>0, \rho >0$. At high temperatures,
the system exhibits algebraic correlations for all $\rho$. 
\newline}
\begin{ruledtabular}
\begin{tabular}{|l|l|l|l|}
 & $\rho=0$ & $\rho>0$ & $\rho<0$ \\
\hline
$T=0$ & KT-like & Plaquette, VBS & Staggered VBS\\
\hline
$T>0$ & No transition & KT & VBS melting \\
\end{tabular}
\end{ruledtabular}
\end{table}

In what follows, we will examine the height model representation of Eq.(\ref{L}) for different values of 
$\rho \propto (t-v)$ and temperatures $T$.
In terms of the effective theory of Eq.(\ref{L}), the RK point corresponds to $\rho=0$. 
For $\rho>0$ the gradient term $(\nabla h)^2$ dominates over 
$(\nabla^2 h)^2$ at large distances, so that the latter can be neglected. 
The free propagator of the height field in momentum space and real time is

\begin{eqnarray}
\label{h-prop}
G_{0} (\omega,p)=  \langle h(\omega,p)h(-\omega,-p)\rangle_{0} \nonumber
\\ =\frac{1}{c\omega^2-\rho p^2-K^{-1}p^4},
\end{eqnarray}
which diverges for an excitation spectrum of the form (\ref{Bog-spec}). As the large distance limit is equivalent to $p\to 0$, we see that 
for $\rho>0$ the $p^4$ term is negligible. By integrating out $h_j$ in Eq. (\ref{Action-0}),
we obtain a charged gas with charges $Q_j=2\pi n_j$ interacting via a potential whose continuum limit 
yields the Euclidean counterpart of Eq. (\ref{h-prop}).  For $K=\infty$, in 
which case the term proportional to $p^4$ is also absent, this is just 
a classical Coulomb gas in three dimensions. It is well known that in this case the excitations 
are gapped,\cite{Polyakov,KNS-2} with a gap that never vanishes. 
This result remains valid for $K$ finite. 
Invoking a 
simple mean-field screening argument, the Debye-H\"uckel approximation, and truncating the 
cosine term in 
(\ref{L}) to lowest (quadratic) order, we arrive at the propagator
\begin{equation}
\label{h-prop-1}
G(\omega,p) =  \langle h(\omega,p)h(-\omega,-p)\rangle= 
\frac{1}{G_{0}^{-1}(\omega,p) - 4 \pi^{2}z}.
\end{equation}
It is readily seen that this propagator screens the interaction between the 
charges of the gas. 
We turn back and ask what phases are
to be found at different values of $\rho$. 

Essentially there are two types of phases. There are valence-bond phases where the 
height field is uniform and others where it modulates 
with some  wave vector $p_0$. The former leads to a height susceptibility that diverges 
for $\omega=0$ and $p=0$, while the latter diverges for $p^2=p_0^2$ when the frequency 
vanishes. In detail, we have the following scenarios:

(a) For $\rho>0$ we can set $K=\infty$, since in this case higher gradient terms are unimportant.  
In this case we have to distinguish between the cases $z<0$ and $z>0$. 
For $z>0$, the height susceptibility (\ref{h-prop-1}) diverges only for 
$\omega=0 $ and $p=0$. This corresponds to the  
plaquette state. \cite{plaq_explain} For $z<0$, on the other hand, there is a divergence for a nonzero wavenumber 
$p_{0}$ given by
\begin{equation}
\label{p0-1}
 p_0^2=-\frac{4\pi^2z}{\rho}.
\end{equation}
This state corresponds to a VBS. 
These two phases --- the plaquette and the VBS (flat) phases --- are schematically shown as 
a function of $\rho$ in Fig \ref{Fig:Plaquette-VBS}. Note that the plaquette phase occurs first, after the RK point ($\rho=0$), before 
the VBS phase; see Refs. \onlinecite{Ralko-2008} and \onlinecite{Leung}.

(b)  For $\rho<0$  the higher gradient term becomes important at large distances as a way to stabilize the system. In this case, the zero-frequency height susceptibility diverges for a nonzero wavenumber
$p_{0}$ given by
\begin{equation}
\label{p0}
 p_0^2=\frac{K}{2}\left(\sqrt{\rho^2-\frac{16\pi^2 z}{K}}-\rho\right).
\end{equation}
This leads to a modulation in the height field 
associated to a staggered VBS state. Note that for $K=\infty$ Eq. (\ref{p0}) reduces to Eq. (\ref{p0-1}), with the difference 
that now it is $\rho$, and not $z$, that is negative.

Below we will describe different regimes in the QDM both at zero and finite temperatures. In what follows it will be more convenient to discuss 
the finite temperature case first, since this essentially corresponds to a classical dimer model.

\subsection{Finite temperature, $\rho>0$}

For $\rho>0$ and high temperatures ($T$), 
we effectively have a classical dimer problem, as in this case $h(\tau,{\bf x})\approx h({\bf x})$ and the system becomes 
effectively two-dimensional. 
\begin{equation}
 S=\int_0^\beta d\tau\int d^2x{\cal L}\approx T^{-1}\int d^2x{\cal L}=\int d^2x{\cal L}_{\rm classical}.
\end{equation}
After some trivial rescaling, we obtain the classical dimer Lagrangian
\begin{equation}
\label{L-cl}
 {\cal L}_{\rm classical}=\frac{1}{2}(\nabla h)^2-\zeta\cos(2\pi\sqrt{\kappa}h),
\end{equation}
with $\kappa=T/\rho$ and $\zeta=z/T$. 
In this limit, we obtain an ordinary two dimensional sine-Gordon theory. \cite{Aletsquare}
This theory describes the vortex unbinding 
transition in the Kosterlitz-Thouless (KT) phase transition.\cite{KT}
It follows that, the high-temperature regime of the RK model for $\rho>0$ undergoes 
{\em a KT phase transition}, 
in which a VBS or a plaquette state \cite{plaq_explain} 
melts into a liquid of valence-bonds, i.e., a finite temperature ``spin liquid''. The Lagrangian (\ref{L-cl}) corresponds to the finite 
temperature dual model \cite{Yaffe} to compact Maxwell electrodynamics in $2+1$ dimensions.\cite{Polyakov} Although at zero temperature 
no phase transition occurs in this model, finite temperature effects lead to 
deconfinement.\cite{Yaffe} In spin models of Mott insulators, a finite temperature deconfinement transition liberates spinon excitations. 
Recently, thermally induced spinon deconfinement 
was shown to occur in a class of pyrochlore antiferromagnets.\cite{Nussinov-2007}
Entropic effects led to a finite temperature phase with algebraic correlations.

We next derive the one-loop effective potential for the classical dimer model. 
Within the height representation, this amounts to the effective potential for the 
two-dimensional sine-Gordon model.\cite{Malik} 
This can be readily acheived by shifting the field $h$ and integrating out the Gaussian 
fluctuations. The effective potential obtained in this way is given by

\begin{eqnarray}
\label{Ueff-finite-T}
&\bar U_{\rm eff}(\varphi)=-\zeta\cos\left(2\pi\sqrt{\kappa}\varphi\right)
\nonumber\\
&-\frac{\pi \kappa\zeta}{2}\cos\left(2\pi\sqrt{\kappa}\varphi\right)
\ln\left[\frac{4\pi^2\kappa\zeta}{\Lambda^2}\cos\left(2\pi\sqrt{\kappa}\varphi\right)\right],
\end{eqnarray}
where $\Lambda$ is an ultraviolet cutoff proportional to the inverse of the lattice spacing. In order to eliminate the dependence on the 
cutoff, we renormalize the above classical effective potential by demanding that $\bar U_{\rm eff}''(0)$ equals $\lambda_D^{-2}=4\pi^2\kappa\zeta$. Here,
$\lambda_D$ is the correlation length associated with the Debye-H\"uckel approximation to (\ref{L-cl}), which consists in making the Gaussian 
approximation $\cos(2\pi\sqrt{\kappa}h)\approx 1-2\pi^2\kappa h^2$. Since 

\begin{equation}
\bar U_{\rm eff}''(0)=4\pi^2\zeta\kappa\left[1+\frac{\pi\kappa}{2}+ \frac{\pi\kappa}{2}\ln\left(\frac{4\pi^2\kappa\zeta}{\Lambda^2}\right)\right],
\end{equation}
we have therefore that the renormalization condition implies $\ln(4\pi^2\kappa\zeta/\Lambda^2)=-1$, such that the effective potential 
becomes

\begin{eqnarray}
\label{Ueff-cl}
&\bar U_{\rm eff}(\varphi)=-\zeta\left(1-\frac{\pi\kappa}{2}\right)\cos\left(2\pi\sqrt{\kappa}\varphi\right)
\nonumber\\
&-\frac{\pi\kappa\zeta}{2}\cos\left(2\pi\sqrt{\kappa}\varphi\right)\ln\left[\cos\left(2\pi\sqrt{\kappa}\varphi\right)\right]. 
\end{eqnarray}
 This minimum energy density 

\begin{equation}
\bar E_0\equiv\bar U_{\rm eff}(0)=\zeta(\pi\kappa/2-1),
\end{equation}
changes sign at $\kappa_c=2/\pi$, i.e., the usual critical value for 
the superfluid stiffness in the KT transition.

That is, the VBS/plaquette system melts from
an ordered crystal to a critical phase 
with algebraic dimer correlations at
the KT temperature
\begin{eqnarray}
T_{KT} = \frac{2}{\pi} \rho.
\label{TKT}
\end{eqnarray}

From Eq. (\ref{Ueff-finite-T}), we see that we can define an effective fugacity given by

\begin{equation}
 \zeta_{\rm eff}=\zeta\left[1+\frac{\pi\kappa}{2}\ln\left(\frac{4\pi^2\kappa\zeta}{\Lambda^2}\right)\right].
\end{equation}
This immediately leads to RG equation for the 
dimensionless coupling $\hat \zeta=\zeta_{\rm eff}/\Lambda^2$, since up to terms of order higher than one loop, we can write 

\begin{equation}
 \Lambda\frac{\partial\zeta_{\rm eff}}{\partial\Lambda}=-\pi\kappa\zeta\approx-\pi\kappa\zeta_{\rm eff},
\end{equation}
so that

\begin{equation}
 \Lambda\frac{\partial\hat \zeta}{\partial\Lambda}=(2-\pi\kappa)\hat \zeta,
\end{equation}
which is precisely the well known flow equation for the fugacity in the KT transition.\cite{KT} 

We now make contact with our relation of Eq.(\ref{TKT})
and a particular limit where its meaning becomes physically transparent.
To this end, we consider the limit of large $\rho$
that physically corresponds to a large negative $v$ for which
the columnar states are the ground states. In this limit, the value of  $T_{KT}$ should  tend
to a divergent value in the limit $ \rho \to \infty$ (or $v \to -\infty$).
Physically,  for $|v| \gg t$ (i.e., ignoring $t$),
the energy cost of breaking the columnar  order
in order to allow for
a liquid state scales with $|v|$ itself.

For $|v| \gg t$ and $v<0$, there are 4 ground states --- the columnar 
ground states. The corresponding phases break a discrete 
symmetry. Similar to the two-dimensional Ising model, \cite{peierls}
these discrete symmetries can and will be broken 
at finite temperatures. Domain wall energy penalties 
(scaling as $|v|$ (or $\rho$) multiplied by the perimeter 
of the domain walls) will, at sufficiently low temperature, overcome
entropic contributions to the free energy
and will force the system to order. 
Similarly,  {\em for any $\rho$} (positive or negative), in the opposite regime of 
infinite temperature, the resultant classical
random dimer covering problem exhibits critical 
correlations.\cite{Fisher} The KT type transition
derived above separates the critical high temperature
and low temperature discrete broken symmetry regimes.
If, unlike the QDM, the Hilbert space would be restricted
to singlet dimers only in a low energy sector and allow
for unconstrained configurations at higher
temperatures,
then the critical phase found would appear only in an intermediate
temperature regime, 
such as may occur in spin systems, e.g., Ref. \onlinecite{Nussinov-2007}.
In such an instance, for high temperatures, the correlations fall off
exponentially. For temperatures in which the 
allowed configurations are restricted to dimer
models, we will find algebraic decay at high
temperatures (as in the infinite temperature
limit of the QDM) and at yet lower temperature,
the system would order into a VBS/plaquette phase
which breaks discrete lattice symmetries. Such
a behavior is found for $Z_{N}$ model with
large, but finite $N$. In such $Z_{N}$ models,
a KT phase is found with both an upper and lower
KT transitions. \cite{FrohSpencer} The behavior of the 
zero temperature QDM, as we will elaborate on 
later, is that of $N \to \infty$.

\subsection{Finite temperature, $\rho=0$}

The case of $\rho=0$ at finite temperature is less obvious. In this case, the finite temperature theory is given by
\begin{equation}
\label{L-cl-1}
 {\cal L}_{\rm classical}=\frac{1}{2\lambda}(\nabla^2 h)^2-\zeta\cos(2\pi h),
\end{equation}
where $\lambda=K/T$. The higher order derivative makes the upper and lower critical dimensions higher than in the standard sine-Gordon theory in 
Eq. (\ref{L-cl}), which has an upper critical dimension $D_{\rm cr}^+=2$. In the case of Eq. (\ref{L-cl-1}) the upper critical dimension is 
$D_{\rm cr}^+=4$, while the lower one is $D_{\rm cr}^-=2$. Therefore, provided the system is at the RK point and finite temperature, {\em no phase 
transition} should occur as the finite temperature system lies effectively at the lower critical dimension, due to dimensional reduction.  In order 
to see more concretely that no phase transition occurs in this case, let us consider the Renormalization
Group flow equations in the general case of $D$ dimensions instead of focusing solely on two dimensions. In principle, a term $\sim (\nabla h)^2$ is generated by fluctuations, but the inclusion of this term simply corresponds to 
a renormalization of $\rho$ and we can 
still demand to be at the RK point by just requiring that the renormalized $\rho$ vanishes. The way of doing such a calculation has been discussed before.\cite{KNS-2} 
The same method can be used here. The resulting flow equations read

\begin{equation}
 \frac{d\lambda^{-1}}{dl}=\zeta^2+(D-4)\lambda^{-1},
\end{equation}

\begin{equation}
\label{flow-zeta}
 \frac{d\zeta}{dl}=\left[D-1-\frac{(D-4)\Gamma(D/2-2)}{8\pi^{D/2-2}}\lambda\right]\zeta.
\end{equation}
We see that the second term between brackets in Eq. (\ref{flow-zeta}) diverges in $D=2$
dimensions. Therefore, the systems does not reach any fixed 
point or line in this case. This more refined analysis clearly shows that no phase transition happens at a finite temperature RK point. 
We note in passing that for $D=4$ the above flow equations lead to a four-dimensional KT transition. Such a transition has been discussed some 
time ago in the context of quantum gravity,\cite{Antoniadis} where a model similar to (\ref{L-cl-1}) was considered. 
 
\subsection{Zero temperature, arbitrary $\rho$}

For $T=0$ and $\rho>0$ there is also {\em no phase transition}. 
As we have already discussed, the higher gradient term becomes in this case RG irrelevant. 
It is well known that the model is in this case equivalent 
via duality to an electromagnetic (Maxwell) theory in $2+1$ dimensions with a {\it compact} gauge field 
\cite{Polyakov} in which the charge-less, 
spin carrying, elementary excitations of the system, the so called spinons, are permanently confined. 
We have described above how spinon deconfinement can arise due to 
thermal effects in the gauge theory.\cite{Yaffe}

\begin{figure}
\includegraphics[width=8cm]{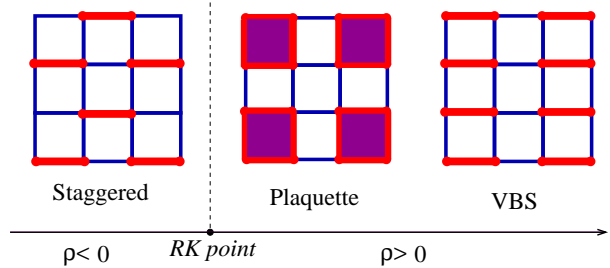}
\caption{(Color online) Schematic representation of the main different phases arising 
in two dimensions and 
at zero temperature. There is numerical evidence for a mixed phase between the VBS and the 
plaquette states (not shown here).\cite{Ralko-2008} The resulting phase transition is discontinuous.}
\label{Fig:Plaquette-VBS}
\end{figure}

Note that the absence of phase transition for $\rho>0$ and $T=0$ follows from an analysis of the effective height model. Thus, the effective model predicts 
a single VBS  or plaquette phase (depending on the sign of $z$) 
 for $\rho>0$ and vanishing temperature. However, it is known that for $\rho>0$ there is a first-order phase transition from the VBS 
state at larger $\rho$ to 
a {\em plaquette} state at smaller $\rho$;\cite{Ralko-2008} 
see Fig. \ref{Fig:Plaquette-VBS}. 
Up to now we have mentioned both phases for $\rho>0$ only in terms of the sign of $z$.  
In order to capture the phase transition, 
a numerical analysis of the lattice gauge theory of Eq. (\ref{Action}) would be necessary. This will be done in a future publication. As a way of making progress by pure analytical means, we can alternatively consider 
higher harmonics of the cosine term in the Lagrangian (\ref{L}) as derived from the action
of Eq.(\ref{Action-0}). This will lead to consider a Lagrangian  that is discussed in Ref. \onlinecite{Ralko-2008}, 
\begin{equation}
\label{L-Ralko}
 {\cal L}=\frac{c}{2}(\partial_\tau h)^2+\frac{\rho}{2}(\nabla h)^2-z_1\cos(2\pi h)
-z_2\cos(4\pi h).
\end{equation}
If both $z_1$ and $z_2$ are positive, then as we discussed earlier the higher cosine harmonic does not play any significant role at large distances, since 
it is RG irrelevant.\cite{ZJ} Thus, in such a situation we have just the plaquette phase as before when 
$z>0$. However, if $z_1<0$ then the term $-z_2\cos(4\pi h)$ is no longer irrelevant if $z_2>0$
and this higher order term adduced from Eq.(\ref{Action-0}) must be kept.

The application of mean-field theory to the Lagrangian
of Eq.(\ref{L-Ralko}) leads to the solution
\begin{equation}
 h_0=\frac{1}{2\pi}\arccos\left(-\frac{z_1}{4z_2}\right).
\end{equation}
By computing the Gaussian fluctuations around $h_0$, we obtain 
a modified Debye-H\"uckel approximation leading to a real-time height susceptibility of the form,
\begin{equation}
 \langle h(\omega,p)h(-\omega,-p)\rangle=
\frac{1}{c\omega^2-\rho p^2-\pi^2z_2^{-1}(z_1^2-16z_2^2)}.
\end{equation}
Note that this type of Debye-H\"uckel approximation is different from the one we have considered before, where the 
height field mean-field was an integer. 
For $z_2>0$ and $(z_1/z_2)^2>16$ we have a gapped spectrum for height field excitations.  When $z_2>0$ and 
$(z_1/z_2)^2<16$, on the other hand, we have a pole for $\omega=0$ at
\begin{equation}
 p_0^2=\frac{\pi^2}{\rho z_2}(16z_2^2-z_1^2).
\end{equation}
These are two distinct regimes in a mixed phase between the 
plaquette and the VBS phases,\cite{Leung} which was recently shown to exist.\cite{Ralko-2008}
Another regime inside the mixed phase is the one 
where $z_1/z_2=\pm 4$, for which the 
excitations are gapless and $h_0=n/2$, with $n\in\mathbb{Z}$, i.e., $h_0$ is either an integer or a 
half odd integer. 
The existence of gapless exitations in the mixed state is important because it demonstrates that there is indeed a phase 
transition between the plaquette and VBS phase. It can now be seen that when $z_2=0$ and $z_1=z$ a simple sign 
change in $z$ does not lead to any phase transition between the plaquette and VBS phases. Indeed, 
after defining the couplings ${\cal K}=(\sqrt{c}\rho)^{-1}$ and $y=\sqrt{c}\rho z$, 
the RG equations would be in this case (the derivation can be easily done with the methods of Ref. \onlinecite{KNS-2})
\begin{equation}
 \frac{dy}{dl}=y\left(3-\frac{{\cal K}}{2}\right),
\end{equation}
\begin{equation}
 \frac{d{\cal K}^{-1}}{dl}={\cal K}^{-1}+y^2.
\label{stiffness-RG}
\end{equation}
Thus, irrespective of the sign of $y$ (or $z$; note that $dy/dl$ also changes sign), the RG equations remain invariant and no 
nontrivial fixed point can be found. The RG approach shows clearly that a sine-Gordon theory in $2+1$ dimensions does not 
exhibit any phase transition. This result is independent of the sign of $z$. This is not the case with the theory of Eq.(\ref{L-Ralko}). 

Let us now turn back to the model with a single cosine harmonic and $K^{-1}\neq 0$, Eq.(\ref{L}), at zero temperature. 
By integrating out the Gaussian fluctuations in the height field 
\begin{widetext}
\begin{equation}
 U_{\rm eff}(\varphi)=-z\cos(2\pi\sqrt{K}\varphi)
+\frac{1}{2}\int\frac{d\omega}{2\pi}\int\frac{d^2p}{(2\pi)^2}\ln\left[\frac{K(c~\omega^2+\rho p^2)+p^4+M_D^2\cos(2\pi\sqrt{K}\varphi)}{K(c~\omega^2+\rho p^2)+p^4}
\right],
\end{equation}
where similarly to the finite temperature case, $M_D^2=4\pi^2Kz$ gives the inverse square of the 
Debye-H\"uckel length.  We obtain
\begin{eqnarray}
 U_{\rm eff}(\varphi)&=&-z\left(1-\frac{\pi}{8}\sqrt{\frac{K}{c}}\right)\cos(2\pi\sqrt{K}\varphi)
-\frac{\rho K}{16}\sqrt{\frac{z}{c}\cos(2\pi\sqrt{K}\varphi)}
\nonumber\\
&+&\sqrt{\frac{K}{c}}\frac{K\rho^2-16\pi^2z\cos(2\pi\sqrt{K}\varphi)}{64\pi}
\ln\left[\frac{K\rho+4\pi\sqrt{Kz\cos(2\pi\sqrt{K}\varphi)}}{4\Lambda^2}\right].
\end{eqnarray}
By using the renormalization condition $U_{\rm eff}''(0)=M_D^2$, we finally obtain the renormalized effective potential
\begin{eqnarray}
\label{Ueff}
 U_{\rm eff}(\varphi)&=&-z\left(1-\frac{\pi}{8}\sqrt{\frac{K}{c}}\right)\cos(2\pi\sqrt{K}\varphi)
-\frac{\rho K}{16}\sqrt{\frac{z}{c}\cos(2\pi\sqrt{K}\varphi)}
\nonumber\\
&+&\sqrt{\frac{K}{c}}\frac{K\rho^2-16\pi^2z\cos(2\pi\sqrt{K}\varphi)}{64\pi}
\ln\left[\frac{K\rho+4\pi\sqrt{Kz\cos(2\pi\sqrt{K}\varphi)}}{K\rho+4\pi\sqrt{Kz}}\right].
\end{eqnarray}
\end{widetext}
At the RK point we have $\rho=0$ and the above effective potential becomes
\begin{eqnarray}
\label{Ueff-RK}
  U_{\rm eff}(\varphi)&=&-z\left(1-\frac{\pi}{8}\sqrt{\frac{K}{c}}\right)\cos(2\pi\sqrt{K}\varphi)
\nonumber\\
&-&\frac{\pi z}{8}\sqrt{\frac{K}{c}}\cos(2\pi\sqrt{K}\varphi)\ln\left[\cos(2\pi\sqrt{K}\varphi)\right].
\nonumber\\
\end{eqnarray}
Interestingly, Eq. (\ref{Ueff-RK}) gives the same type of effective potential as in Eq. (\ref{Ueff-cl}), which was obtained for $\rho>0$ and 
high temperatures. 
Note, however, the presence of the extra parameter $c$ in Eq. (\ref{Ueff-RK}). 
On the other hand, if we introduce the dimensionless parameters $\hat c=c/\Lambda$ and $\hat K=K/\Lambda$ along with $\phi=\sqrt{\Lambda}\varphi$, we 
see that for the special choice $\hat c=1/\hat K$ the effective potential becomes 
\begin{eqnarray}
\label{Ueff-RK-1}
  U_{\rm eff}(\phi)&=&-z\left(1-\frac{\pi\hat K}{8}\right)\cos(2\pi\sqrt{\hat K}\phi)
\nonumber\\
&-&\frac{\pi z\hat K}{8}\cos(2\pi\sqrt{\hat K}\phi)\ln\left[\cos(2\pi\sqrt{\hat K}\phi)\right],
\nonumber\\
\end{eqnarray}
which has now exactly the same form as (\ref{Ueff-cl}).  
The above effective potential corresponds to a {\em KT-like transition with a critical stiffness $\hat K_c=8/\pi$}. Indeed, the minimum 
energy density is given by

\begin{equation}
 E_0\equiv U_{\rm eff}(0)=-z\left(1-\frac{\pi\hat K}{8}\right),
\end{equation}
which changes sign precisely at $\hat K_c=8/\pi$. 
A KT-like transition at zero temperature was also obtained recently for a QDM.\cite{Castelnovo,Ardonne} 

There is a deeper argument behind the choice $\hat c=1/\hat K$. It actually corresponds to a self-dual point of the model in a special 
limit. To see this, first we note that for $\rho=0$ the Lagrangian (\ref{L}) is dual to the U(1) symmetric lattice Lagrangian  
\cite{Amit-1982}
\begin{equation}
\label{U(1)}
 {\cal L}_{U(1)}=\sum_i\left[\frac{\hat K}{2}(\nabla_\tau\theta_i-2\pi n_i)^2+\frac{1}{2\hat c}(\nabla^2 \theta_i-2\pi m_i)^2\right],
\end{equation}
where $n_i,m_i\in\mathbb{Z}$ and $-\pi<\theta_i\le \pi$. The self-duality and its relation to a KT transition in $2+1$ dimensions follows 
from the dualization of the following $Z_N$ model, 
\begin{eqnarray}
\label{Z_N}
 {\cal L}_{Z_N}&=&\sum_i\left[\frac{\hat K}{2}\left(\frac{2\pi}{N}\nabla_\tau q_i-2\pi n_i\right)^2\right.
\nonumber\\
&+&\left.\frac{1}{2\hat c}\left(\frac{2\pi}{N}\nabla^2 q_i-2\pi m_i\right)^2\right],
\end{eqnarray}
in the $N\to\infty$ limit, 
where $q_i=0,1,2,\dots,N-1$. 
In the limit $N\to\infty$ the Lagrangian (\ref{Z_N}) becomes (\ref{U(1)}). 
As shown by Amit {\it et al.},\cite{Amit-1982} a duality transformation relates the 
partition function of the $Z_N$ model to its dual as 
\begin{equation}
Z(\hat K,1/\hat c)\sim Z(N^2\hat c/(4\pi^2),N^2/(4\pi^2\hat K)),
\end{equation}
and we see that the couplings are not only inverted by duality, but also exchanged. Therefore, we obtain the self-duality relation 
$\hat c/\hat K=4\pi^2/N^2$. For large $N$ the self-dual point with $\hat c=1/\hat K$ leads to a phase transition as $\hat K$ is varied. 
This self-duality point is precisely what we have demanded to obtain Eq. (\ref{Ueff-RK-1}), with a resulting KT transition. 
The self-duality breaks down along the line $\hat c=\hat K$. In this case no phase transition occurs. 

Note that in spite of the similarities between the described self-dual quantum critical point and the usual KT transition at finite 
temperature, there are actually important differences, making them distinct universality classes. That is the reason why we refer to 
the self-dual quantum critical point as ``KT-like'' rather than simply ``KT''. In order to emphasize this point, let us consider 
the spin-wave theory of the the model (\ref{U(1)}) in the continuum limit and at the self-dual point. We have the correlation function between 
the $XY$ spins:

\begin{eqnarray}
 C(\tau,{\bf r})&=&\langle e^{i\theta(\tau,{\bf r})}e^{-i\theta(0,0)}\rangle
\nonumber\\
&=&{\cal N}\exp\left\{\frac{1}{\hat K}\left[G(\tau,{\bf r})-G(0,0)\right]\right\},
\end{eqnarray}
where ${\cal N}$ is a normalization constant and

\begin{equation}
 G(\tau,{\bf r})=\int\frac{d\omega}{2\pi}\int\frac{d^2p}{(2\pi)^2}\frac{e^{i(\omega\tau-{\bf p}\cdot{\bf r})}}{\omega^2+p^4}.
\end{equation}
In the static limit we have a regime looking very much similar to KT, since

\begin{equation}
 G(0,{\bf r})\approx-\frac{1}{4\pi}\ln\left(\frac{r}{4a}\right)+{\rm const},
\end{equation}
where $a$ is a cutoff. However, this leads to a different anomalous dimension for the correlation function $C(\tau,{\bf r})$ in the 
limit $\tau\to 0$, i.e., 

\begin{equation}
 \eta=\frac{1}{4\pi\hat K_c}=\frac{1}{32},
\end{equation}
which is to be compared to $\eta=1/4$ for the usual KT transition.

On the other hand, note that for large $\tau$ the Green function behaves like

\begin{equation}
 G(\tau,{\bf r})\approx \frac{a^2}{4\pi\tau}\exp\left(-\frac{a^2r^2}{2\tau^2}\right),
\end{equation}
which can be derived from the exact representation

\begin{eqnarray}
 G(\tau,{\bf r})&=&\frac{1}{4\pi}\int_0^{a^2}\frac{du}{\sqrt{\tau^2+4u^2}}I_0\left(\frac{ur^2}{2(\tau^2+4u^2)}\right)
\nonumber\\
&\times&\exp\left[-\frac{ur^2}{2(\tau^2+4u^2)}\right],
\end{eqnarray}
where $I_0$ is a modified Bessel function of the first kind. For large distances with {\it finite} $\tau$ we have an asymptotically static 
behavior:

\begin{equation}
 G(\tau,{\bf r})\approx\frac{a}{2\pi^{3/2}r}.
\end{equation}

\begin{figure}
\includegraphics[width=8cm]{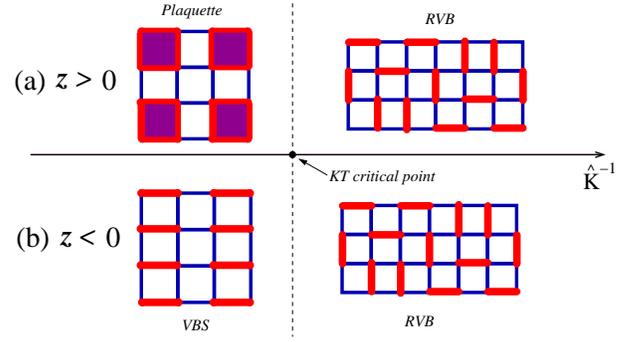}
\caption{(Color online) Schematic representation of the phases of the two-dimensional QDM 
at the RK point.  The zero temperature 
KT (quantum) critical point is given by $\hat K_c=8/\pi$.
Note that for $z>0$ [Panel (a)], the KT-like transition is from a 
plaquette to an RVB state, which is a liquid of valence bonds. In Panel (b) the RVB state emerges out of a VBS state at 
$z<0$.}
\label{Fig:RK-phases}
\end{figure}

The zero temperature transition of the two-dimensional QDM describes a phase transition where a plaquette or VBS state changes into a spin liquid, 
here thought as a liquid of valence bonds. A very schematic representation of the phase transition at zero temperature is 
shown in Fig. \ref{Fig:RK-phases}.  
As in the finite temperature KT transition of the QDM, we have here that a VBS state corresponds to a ``superfluid'' phase, while the 
spin liquid will be the ``normal fluid'' of the transition. 

\begin{figure}
\includegraphics[width=8cm]{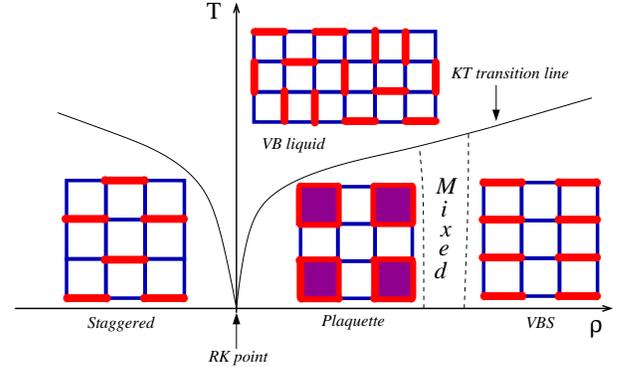}
\caption{(Color online) Schematic phase diagram of the two-dimensional QDM 
in the $\rho T$-plane.  For large $\rho$,
the KT temperature is high and varies linearly with $\rho$ with a slope of $(2/\pi)$ (Eq.(\ref{TKT})).
Here we have also indicated the mixed phase.\cite{Ralko-2008,Leung} 
Temperature effects melt the dimerized phases to a valence-bond (VB) liquid phase, which is actually the same 
as a thermally induced RVB state.}
\label{Fig:RK-phase-diag-rho-T}
\end{figure}

Putting all of the pieces together, we arrive at Table I that encapsulated our 
results at the beginning of this section.  We obtained these results 
from the analysis of effective height models via duality arguments and the study of effective
potentials. A schematic phase diagram in the $\rho T$-plane is shown in Fig. \ref{Fig:RK-phase-diag-rho-T}.


\section{Three-dimensional effective model for the RK point and duality}
\label{3Dsection}

Now we will consider the $(3+1)$-dimensional quantum dimer model. 
In this case no pure scalar field theory arises via duality. 
The effective model in this case is a 
compact gauge theory,\cite{Review,Fradkin-2004}
\begin{equation}
\label{QDM-3D}
 {\cal L}=\frac{c}{2}(\partial_\tau{\bf A})^2+\frac{\rho}{2}(\nabla\times{\bf A})^2+\frac{1}{2K}(\nabla\times\nabla\times{\bf A})^2.
\end{equation}
As before, the RK point corresponds to $\rho=0$. 
We now derive the corresponding dual model at the RK point. This is easily done by considering the lattice model version 
of (\ref{QDM-3D}) in the Villain approximation. In this approximation, 
the lattice action reads

\begin{eqnarray}
\label{S-QDM-3d-RK}
 S&=&\sum_j\left[\frac{c}{2}(\nabla_\tau{\bf A}_j-2\pi{\bf n}_j)^2\right.
\nonumber\\
&+&\left.\frac{1}{2K}(\nabla\times\nabla\times{\bf A}_j-2\pi{\bf m}_j)^2\right],
\end{eqnarray}
where ${\bf n}_j$ and ${\bf m}_j$ are integer fields and the lattice sites $j$ are in $3+1$ dimensions. 
The first important step is to introduce new integer-valued auxiliary fields 
using the Poisson formula. This leads to the action
\begin{eqnarray}
 S&=&\sum_j\left[\frac{1}{2c}{\bf M}^2_j+\frac{K}{2}{\bf N}^2_j-i{\bf M}_j\cdot\nabla_\tau{\bf A}_j\right.\nonumber\\
&-&\left.i{\bf N}_j\cdot\nabla\times\nabla\times{\bf A}_j\right],
\end{eqnarray}
where ${\bf M}_j$ and ${\bf N}_j$ are integer fields.
As the gauge field ${\bf A}_j$ is compact, integrating it out   
yields the local constraint 

\begin{equation}
\nabla_\tau{\bf M}_j+\nabla^2{\bf N}_j-\nabla(\nabla\cdot{\bf N}_j)=0.
\end{equation}
This constraint 
is solved by the equations 
\begin{equation}
{\bf M}_j=\nabla\times\nabla\times{\bf L}_j,~~~~~~~~~{\bf N}_j=\nabla_\tau{\bf L}_j.
\end{equation}
Invoking the Poisson 
formula once more to promote the integer field ${\bf L}_j$ to a real field ${\bf a}_j$, we obtain the action of the dual model in 
the form
\begin{equation}
\label{S-dual-3d}
 \tilde S=\sum_j\left[\frac{K}{2}(\nabla_\tau{\bf a}_j)^2+\frac{1}{2c}(\nabla\times\nabla\times{\bf a}_j)^2-2\pi i{\bf n}_j\cdot{\bf a}_j\right],
\end{equation}
where ${\bf n}_j$ is an integer-valued field. Two important remarks are in order. First, we note that, as before, the couplings are 
inverted and exchanged. However, {\it the gauge field appearing in the dual model is not compact}. Second, there is a coupling between the gauge field and 
an integer field. To understand the physical meaning of this coupling, note that due to gauge invariance 
in space direction \cite{Note-1} the constraint ${\nabla}\cdot{\bf n}_j=0$ holds. 
This means that we can sum over field configurations with loops (akin to vortex loops).\cite{Kleinert-Book,Dasgupta,Kleinert-Tric}
This constraint enters in the partition function as a Kronecker delta. Employing the integral representation of the Kronecker delta and partial 
summation yields
\begin{eqnarray}
 \tilde S&=&\sum_j\left[\frac{K}{2}(\nabla_\tau{\bf a}_j)^2+\frac{1}{2c}(\nabla\times\nabla\times{\bf a}_j)^2\right.\nonumber\\
&+&\left.i(\nabla\theta_j-2\pi{\bf a}_j)\cdot{\bf n}_j\right],
\end{eqnarray}
where $\theta_j\in[0,2\pi]$. In the sum over ${\bf n}_j$, we introduce in the action
a soft vortex-core term $\epsilon_c{\bf n}_j^2/2$ 
with a small core energy $\epsilon_c$.\cite{Peskin} By the Poisson summation formula,
\begin{eqnarray}
\label{dual-QDM-3D}
 \tilde S&=&\sum_j\left[\frac{K}{2}(\nabla_\tau{\bf a}_j)^2+\frac{1}{2c}(\nabla\times\nabla\times{\bf a}_j)^2\right.\nonumber\\
&+&\left.\frac{1}{2\epsilon_c}(\nabla\theta_j-2\pi{\bf a}_j-2\pi{\bf m}_j)^2\right].
\end{eqnarray} 

In the continuum limit, the dual model (\ref{dual-QDM-3D}) reads
\begin{eqnarray}
\label{dual-L}
 \tilde {\cal L}&=&\frac{K}{2}(\partial_\tau{\bf a})^2+\frac{1}{2c}(\nabla\times\nabla\times{\bf a})^2
\nonumber\\
&+&|(\nabla-2\pi i{\bf a})\psi|^2+r|\psi|^2+\frac{u}{2}|\psi|^4,
\end{eqnarray}
where $\psi$ is a complex scalar field. 
Quantum fluctuations generate a term $\sim\ln(\xi\Lambda)(\nabla\times{\bf a})^2$
(with a 
{\it positive} coefficient), where 
$\xi$ is the correlation length. This term renders the double-curl term irrelevant at 
large distances. Quantum fluctuations also generate a $|\partial_\tau\psi|^2$ term. 
The resulting effective action is similar to that of an Abelian Higgs model in four spacetime dimensions. The renormalization 
group (RG) flow for such a theory is known \cite{CW} to exhibit a Gaussian 
fixed point with a runaway flow. The 
same occurs here and no nontrivial fixed point is found. Indeed, the $|\psi|^4$ coupling is marginally irrelevant. This leads to a first-order phase transition. Thus, no quantum critical point
appears in this case. 

An interesting aspect of this first-order phase transition is that the mass gap of the gauge field is induced by fluctuations, through the 
so--called Coleman-Weinberg mechanism.\cite{CW,Note-2} 
Here we have an interesting modified version of it. 
This can be seen more straightforwardly by integrating out the gauge field while neglecting 
the scalar field fluctuations, i.e., assuming $\psi$ uniform. This yields the effective potential,

\begin{eqnarray}
\label{Ueff-CW}
U_{\rm eff}(\psi)&=&\left(r+\frac{\pi}{8}\sqrt{\frac{c}{K}}\right)|\psi|^2+\frac{u}{2}|\psi|^4
\nonumber\\
&-&\frac{\pi}{8}\sqrt{\frac{c}{K}}|\psi|^2\ln\left(\frac{\pi^2c|\psi|^2}{\Lambda^2}\right),
\end{eqnarray}
where $\Lambda$ is an ultraviolet cutoff. Note that, in contrast with Ref. \onlinecite{CW}, there is a 
prefactor of $|\psi|^2$ instead of a prefactor $|\psi|^4$ 
that multiplies the logarithmic term. This is a consequence of the $p^4$ term 
in the gauge field propagator. 

The first-order phase transition disorders a three-dimensional VBS into a RVB state. Recall that in two spatial dimensions the quantum phase transition 
from a VBS to a RVB state exhibited a quantum critical point governed by a KT transition. 

It is also useful to consider the case $\rho>0$ and $K \to \infty$, in which case we have instead Eq. (\ref{S-QDM-3d-RK}) the lattice action, 
\begin{equation}
\label{S-QDM-3d}
 S=\sum_j\left[\frac{c}{2}(\nabla_\tau{\bf A}_j-2\pi{\bf n}_j)^2
+\frac{\rho}{2}(\nabla\times{\bf A}_j-2\pi{\bf m}_j)^2\right].
\end{equation}
The first step of the duality transformation leads in this case to an action featuring integer fields ${\bf M}_j$ and ${\bf N}_j$ of the form
\begin{equation}
 S=\frac{1}{2}\sum_j\left(\frac{1}{c}{\bf M}_j^2+\frac{1}{\rho}{\bf N}_j^2\right),
\end{equation}
where the constraint
\begin{equation}
 \nabla_\tau{\bf M}_j=\nabla\times{\bf N}_j,
\end{equation}
must be fulfilled. The constraint is easily solved by introducing a new 
integer field ${\bf L}_j$ such that ${\bf M}_j=\nabla\times{\bf L}_j$ and ${\bf N}_j=\nabla_\tau{\bf L}_j$. With the help of the Poisson formula we 
obtain finally

\begin{equation}
 S=\sum_j\left[\frac{1}{2}(\nabla_\tau{\bf a}_j)^2+\frac{\rho}{2c}(\nabla\times{\bf a}_j)^2-2\pi i\sqrt{\rho}{\bf n}_j\cdot{\bf a}_j\right].
\end{equation}
Here the field ${\bf a}_j$ is a real field while ${\bf n}_j$ is an integer field and we have performed a trivial rescaling. The above action 
is similar to the action (\ref{S-dual-3d}), except that we have here a normal curl term instead of a double-curl one. Therefore, the continuum limit of the dual lattice model above reads
\begin{eqnarray}
\label{dual-L+}
 \tilde {\cal L}_{\rho>0}&=&\frac{1}{2}(\partial_\tau{\bf a})^2+\frac{\rho}{2c}(\nabla\times{\bf a})^2
\nonumber\\
&+&|(\nabla-2\pi i\sqrt{\rho} ~{\bf a})\psi|^2+r|\psi|^2+\frac{u}{2}|\psi|^4.
\end{eqnarray}
At zero temperature the above Lagrangian describes effectively the same first-order phase 
transition as in the case of the theory (\ref{dual-L}) featuring a double-curl. Therefore, 
we see that for $\rho\geq 0$ no quantum critical point exists. The first-order phase 
transition creates a three-dimensional liquid of valence bonds out of a VBS state. 

At high 
temperatures, on the other hand, the situation is very different, 
since in this case ${\bf a}(\tau,{\bf x})\approx{\bf a}({\bf x})$. 
and  the term $(\partial_\tau{\bf a})^2$ can be neglected. 
By defining new fields according to $\Psi=\psi/\sqrt{T}$ and ${\bf h}={\bf a}\sqrt{\rho/(cT)}$, we obtain
\begin{equation}
\label{GL-model}
 \tilde {\cal L}_{\rho>0}\approx\frac{1}{2}(\nabla\times{\bf h})^2+|(\nabla-ie{\bf h})\Psi|^2+r|\Psi|^2+\frac{u}{2}|\Psi|^4,
\end{equation}
where $e=2\pi\sqrt{cT}$ and $v=uT$. The above expression is exactly the same as 
the Ginzburg-Landau (GL) free energy of a superconductor in three dimensions.
In this case, it is known that the order of the phase 
transition is determined by the ratio $w= v/(2e^2)$.\cite{Kleinert-Book,Kleinert-Tric,Sudbo-Tric} 
For $w>0.55$, the phase transition is second-order,\cite{Kleinert-Tric,Sudbo-Tric} 
otherwise a first-order transition takes place. Duality arguments and Monte Carlo simulations 
\cite{Dasgupta,Olsson} indicate that when $w>0.55$, the 
correlation length and specific heat exponents attain XY values, otherwise the phase 
transition is a first-order one. 

At high temperatures a similar analysis can be made 
in the dual model (\ref{dual-L}) to obtain the Lagrangian  
\begin{equation}
\label{GL-2curl}
 \tilde {\cal L}\approx\frac{1}{2}(\nabla\times\nabla\times{\bf h})^2+|(\nabla-ie{\bf h})\Psi|^2+r|\Psi|^2+\frac{v}{2}|\Psi|^4,
\end{equation}
where, as before, $e=2\pi\sqrt{cT}$ and $v=uT$. If it was not for the double-curl term, the above expression would be exactly the same as 
the free energy for a superconductor in three dimensions. 
However, we can proceed similarly as in the zero temperature case and argue 
that thermal fluctuations will generate a $(\nabla\times{\bf h})^2$ term with a positive coefficient, making once more the double-curl term 
irrelevant. Note that this time the generated curl term {\it does not} contain any logarithm in 
front of it, since the high temperature theory is three-dimensional. 
Thus, by taking thermal fluctuations into account, the finite temperature theory 
behaves in the same way as a GL superconductor in 
three dimensions. This result approximately agrees 
with recent simulations of a classical dimer model.\cite{Charrier}  However, recent results suggest a different universality class.
\cite{Misguich-2008-PRB,Powell} In fact, our argument of irrelevance of 
the double-curl has to be taken with care. In order to appreciate the subtleties of the problem, 
let us consider a simple one-loop calculation. In this case, the generated single-curl term 
is $\sim \xi e^2(\nabla\times{\bf h})^2$, with a correlation length $\xi\sim(r-r_c)^{-\nu}$, 
where $r_c$ is the value of $r$ at the critical point and $\nu$ is the corresponding critical 
exponent. Thus, it seems that for fixed $e^2$ and very near to $r_c$, the coefficient of the 
generated $(\nabla\times{\bf h})^2$ term is large. Therefore, for large distances (corresponding to 
$|{\bf p}|\to 0$) the double-curl term seems to be indeed irrelevant. However, this argument may 
be too simplistic for the following reason. Simple dimensional analysis of the Lagrangian 
(\ref{GL-2curl}) shows that $e^2$ has dimension of (length)$^{-3}$. Thus, its renormalized 
counterpart should behave near the critical point as $e_R^2\sim\xi^{-3}$ so that the actual 
one-loop renormalized coefficient of the fluctuation-generated $(\nabla\times{\bf h})^2$ term 
behaves, after replacing $e^2$ by $e_R^2$, as $\sim\xi^{-2}$. Therefore, precisely at the critical point this term vanishes and 
the double-curl term plays again a role. 
However, if we start with the theory already at the critical point, i.e., with $r=r_c$, it is 
easy to determine the anomalous dimension $\eta_h$ of ${\bf h}$ to all orders in peturbation theory, 
provided a perturbative fixed point is found. The method resembles the one used in the 
study of critical fluctuations in superconductors.\cite{Herbut} In the present case it yields 
$\eta_h=3$, such that the gauge field propagator has the behavior

\begin{eqnarray}
 \langle h_i({\bf p})h_j(-{\bf p})\rangle&\sim&\frac{1}{|{\bf p}|^{4-\eta_h}}
\left(\delta_{ij}-\frac{p_ip_j}{{\bf p}^2}\right)\nonumber\\
&\sim&\frac{1}{|{\bf p}|}
\left(\delta_{ij}-\frac{p_ip_j}{{\bf p}^2}\right),
\end{eqnarray}
which is identical to the infrared behavior of the GL model.\cite{Herbut}
This highlights, once more, the irrelevance of the the double-curl term at large distances. 
Thus, we see that there are two possible situations relative the critical point, 
that correspond to two different fixed points. 
This rough analysis shows that the 
double-curl term in Eq. (\ref{GL-2curl}) is dangerously irrelevant. 
Thus, in order to unambiguously determine the universality class of the model of Eq.(\ref{GL-2curl}),   
a more quantitative analysis is necessary. This will be subject of a future publication.  

While in the case of the Lagrangian of Eq. (\ref{GL-2curl}) the dangerously irrelevance of the 
double-term makes it questionable to discard it at the critical point, the same problem does 
not arise with Eq. (\ref{QDM-3D}) for a fixed $\rho>0$. In that case the double-curl term can safely 
be neglected at large distances and the lattice model (\ref{S-QDM-3d}) 
corresponding to $K \to \infty$ can be used. 

\section{Conclusion}
\label{endsection}

In this paper we have derived an extended lattice gauge theory action for the QDM. From it we were able to derive effective models to study 
the phase structure of two- and three-dimensional QDMs. For the two-dimensional 
case we have calculated the one-loop effective potential. At the RK point the effective potential is characteristic of one exhibiting a 
line of fixed points leading to a KT-like transition. This KT-like transition in $2+1$ dimensions and zero temperature is a consequence of the 
anisotropy and self-duality of the effective model. 
It governs a universality class which is actually distinct from the traditional KT transition. 
This is also in stark contrast with the $(2+1)$-dimensional KT-like transition discussed in 
the past in the context of relativistic $U(1)$ spin liquids.\cite{U(1)-spin-liquid} In that case the system was isotropic and the 
propagator non-analytic. It is still controversial whether a $(2+1)$-dimensional KT transition may occur in this case.\cite{Nikolaou}

The quantum critical point of the two-dimensional QDM was related to a self-dual point of a family of 
$Z_N$ theories in the limit $N\to\infty$. This self-duality is strongly dependent of being at the RK point. One interesting aspect in the 
duality scenario of the two-dimensional QDM is that the RK point is protected by the duality symmetry. In other words, as a 
consequence of the intrinsic anisotropy, the parameter 
$\rho\propto (t-v)$ does not get inverted by the duality transformation, so that the RK point (i.e., the point for which $\rho=0$) is 
preserved by the duality transformation. 

It is well known \cite{FrohSpencer} that in $1+1$ dimensions the standard $Z_N$ model (i.e., without higher gradients) undergoes a KT transition for 
$N$ large. This is not surprising, since the $Z_N$ model becomes for $N\to\infty$ the $XY$ model. In our case, the KT transition occurs one dimension higher as a
consequence of the higher gradient term and the anisotropy. To this KT transition there is also a corresponding $Z_N$ model, Eq. (\ref{Z_N}), whose large 
$N$ critical point governs the KT transition of the height model at the RK point. As we have discussed, in this case there is an interesting additional feature, 
which is the self-duality at large $N$.  

At finite temperature and above the RK point ($t>v$) a genuine KT transition occurs. In contrast with the zero temperature KT-like transition, the 
finite temperature transition is a consequence of dimensional reduction 
induced by temperature 
rather than of the intrinsic anisotropy of the system. At the RK point, on the 
other hand, no phase transition occurs at finite temperature. In this case 
dangerous infrared singularities prevent a phase transition to happen. 

For the three-dimensional case we have derived the dual model at the RK point and shown that at 
zero temperature a first-order phase transition takes place. A second-order phase transition can only occur in this case at finite temperature. The finite temperature phase diagram features first- and 
second-order phase transitions separated by a tricritical point, in a scenario very reminiscent from 
phase transitions in superconductors.\cite{Kleinert-Book,Kleinert-Tric,Sudbo-Tric} 

Finally, it should be noted that the approach employed in this paper can be used to derive far more accurate results than those that we have discussed. 
In particular, the phase boundaries near the RK point in the schematic phase diagram can in principle be determined more precisely via Monte Carlo simulations 
of the extended lattice gauge theory introduced in Sect. III. This will be the subject of a future publication.

\acknowledgments
F.S.N. would like to thank the partial support of the Deutsche Forschungsgemeinschaft (DFG), grant No. KL 256/46-1. Z.N. would like to acknowledge the CMI of 
WU.


\begin{thebibliography}{100}
\bibitem{plaq_explain}
The ideal plaquette state is given by
\begin{eqnarray}
|\psi_{plaq} \rangle = \prod_{\Box \in \mbox{sublattice}} \frac{1}{\sqrt{2}} (|\vv\rangle + |\hh\rangle),
\label{ideal_plaq}
\end{eqnarray}
with the roduct over all plaquettes belonging to one of the sublattices
of plaquettes of the square lattice. (There are 4 such sublattices.)
An order parameter \cite{Syljuasen, Leung} 
to detect columnar order/plaquette order 
on a square latice of size $L \times L$ is 
given by $M_{vh} = \frac{2}{L^{2}}(N_{v}- N_{h})$
with $N_{v,h}$ the number of horizontal/vertical dimers. The ideal
columnar state has $M_{v,h}^{2}=1$ while within the ideal plaquette
state of Eq.(\ref{ideal_plaq}) $M= \frac{4}{L^{2}} \to 0$
as the system size tends to infinity.
\end{thebibliography}
\end{document}